\documentclass[showpacs,twocolumn,aps,nofootinbib,nobibnotes]{revtex4}%
\usepackage{amstext}
\usepackage{amsmath}
\usepackage{amssymb}
\usepackage{graphicx}
\usepackage{latexsym}
\usepackage{bm}
\usepackage{amsfonts}%
\begin{document}
\title{Analogue to multiple electromagnetically induced transparency in all-optical
drop-filter systems}
\author{Yun-Feng Xiao\footnote{Present address: Department of Electrical and Systems
Engineering, Washington University in St. Louis, St. Louis, Missouri
63130-4899; electronic mail: yfxiao@gmail.com}}
\author{Xu-Bo Zou}
\email{xbz@ustc.edu.cn}
\author{Wei Jiang}
\author{You-Ling Chen}
\author{Guang-Can Guo}
\affiliation{Key Laboratory of Quantum Information, University of Science and Technology of
China, Hefei 230026, China.}

\begin{abstract}
We theoretically study a parallel optical configuration which includes $N$
periodically coupled whispering-gallery-mode resonators. The model shows an
obvious effect which has a direct analogy with the phenomenon of multiple
electromagnetically induced transparency in quantum systems. The numerical
simulations illuminate that the frequency transparency windows are sharp and
highly transparent. We also briefly discuss the experimental feasibility of
the current scheme in two practical systems, microrings and microdisks.

\end{abstract}

\volumeyear{year}
\volumenumber{number}
\issuenumber{number}
\eid{identifier}
\maketitle

Electromagnetically-induced transparency (EIT) \cite{Harris}, which is based
on the destructive quantum interference, is an interesting phenomenon that the
absorption of a probe-laser field which is resonant with an atomic transition
can be reduced or even eliminated by applying a strong driving laser beam at a
different frequency. Since experimental observation in atomic vapors
\cite{observation}, this effect is playing an essential role in a variety of
physical processes, ranging from lasing without inversion \cite{Zibrov},
enhanced nonlinear optics \cite{Imamoglu} to quantum computation and
communication \cite{Lukin}.

Recent theoretical analysis of optical coupled resonators (or cavities)
without the use of atomic resonance have revealed that coherent effects in
coupled resonator systems are remarkably similar to those in atoms. In Ref.
\cite{smith}, it was shown that the EIT-like effect can be established in
directly coupled optical resonators due to mode splitting and classical
destructive interference \cite{smith}. In Ref. \cite{fan}, it was pointed out
that the existence of a classical analogue of the electromagnetically induced
transparency in coupled optical resonators is crucial for on-chip coherent
manipulation of light at room temperatures, including the capabilities of
stopping, storing and time reversing an incident optical pulse. More recently,
some experiments have been reported for observing the structure tuning of the
EIT-like spectrum in a compound glass waveguide platform using relatively
large resonators \cite{chu}, coupled fused-silica microspheres
\cite{naweed,kouki} and integrated micron-size silicon optical resonator
systems \cite{xu}. These experiments open up the new possibility for optical
communication and simulation of coherent effect in quantum optics using
multiple coupled optical resonators \cite{poon}.

In this paper, we study multiple EIT-like transmission spectrum by means of
$N$ indirectly coupled resonators (via two parallel waveguides, namely, bus
and drop) which is a generalization of the two coupled resonators system
\cite{maleki}. $N$ indirectly coupled resonators result in $N-1$ frequency
transparency windows with equal separation. The numerical simulations indicate
that these frequency windows are ultra-sharp and highly transparent with
practical parameters. The experimental feasibility of the current scheme in
two practical systems, microrings and microdisks, will be finally discussed briefly.

\begin{figure}[pb]
\centerline{\includegraphics[keepaspectratio=true,width=0.5\textwidth]{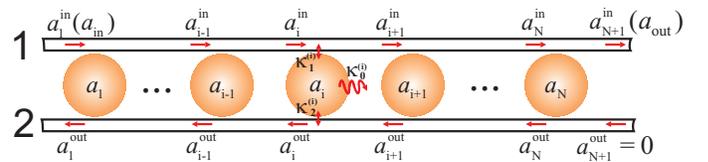}}\caption{(Color
online) Coupled microcavity resonators sharing two waveguides 1 and 2, namely,
bus and drop. As a run-of-mill example, loss parameters in cavity i are
detailedly described.}%
\end{figure}

Consider an array of resonators in which each resonator is coupled to the
adjacent resonators by means of two parallel waveguides, as shown in Fig. 1.
The cavities are labeled in ascending order from left to right, and it is
assumed that there are $N$ cavities altogether. In the case of slowly varying
field amplitudes, the modes of this system can be described by the coupled
harmonic oscillator model. The motion of the i-th ($\mathrm{i=1,2,...,N}$)
cavity mode $a_{\text{\textrm{i}}}$\ (with the center frequency $\omega
_{\mathrm{i}}$) is
\begin{align}
\frac{da_{\text{\textrm{i}}}}{dt}  &  =i\left(  \omega-\omega_{\mathrm{i}%
}\right)  a_{\mathrm{i}}-\frac{\kappa_{\mathrm{0}}^{\mathrm{(i)}}%
+\kappa_{\mathrm{1}}^{\mathrm{(i)}}+\kappa_{\mathrm{2}}^{\mathrm{(i)}}}%
{2}a_{\mathrm{i}}\nonumber\\
&  -\sqrt{\kappa_{\mathrm{1}}^{\mathrm{(i)}}}a_{\mathrm{i}}^{\mathrm{in}%
}-\sqrt{\kappa_{\mathrm{2}}^{\mathrm{(i)}}}e^{i\phi_{\mathrm{i}}%
}a_{\mathrm{i+1}}^{\mathrm{out}}. \label{initial}%
\end{align}
Here $\omega$ denotes the carrier frequency of the input laser field;
$\kappa_{\mathrm{0}}^{\mathrm{(i)}}$, $\kappa_{\mathrm{1}}^{\mathrm{(i)}}$ and
$\kappa_{\mathrm{2}}^{\mathrm{(i)}}$ represent the linewidth associated with
the intrinsic cavity losses, the coupling to waveguides 1 and 2, respectively;
$a_{\mathrm{i}}^{\mathrm{in}}$ and $a_{\mathrm{i+1}}^{\mathrm{out}}$ describe
the input fields of the \textrm{i}-th cavity from waveguides 1 and 2,
respectively, as shown in Fig. 1. The two output fields (transmission and
reflection) are related to the input fields by the output-input relations
$a_{\mathrm{i}}^{\mathrm{in}}=\exp\left(  i\phi_{\mathrm{i-1}}\right)  \left(
a_{\mathrm{i-1}}^{\mathrm{in}}+\sqrt{\kappa_{\mathrm{1}}^{\mathrm{(i)}}%
}a_{\mathrm{i-1}}\right)  $ and $a_{\mathrm{i}}^{\mathrm{out}}=\exp\left(
i\phi_{\mathrm{i}}\right)  a_{\mathrm{i+1}}^{\mathrm{out}}+\sqrt
{\kappa_{\mathrm{2}}^{\mathrm{(i)}}}a_{\mathrm{i}}$. Here $\phi_{\mathrm{i}}$
stands for the phase delay along the waveguides between the adjacent cavities
\textrm{i} and \textrm{i+1}. For simplicity and convenience, we put
$\exp\left(  i\phi_{\mathrm{i}}\right)  =1$ by choosing appropriate
resonator-resonator distance $L$.

We are interested in the steady-state regime of the current system. To
facilitate the discussion, we also suppose that each cavity has the same
dissipation, which originate from the intrinsic loss and the two waveguides,
i.e., $\kappa_{\mathrm{0}}^{\mathrm{(i)}}=\kappa_{\mathrm{0}}$, $\kappa
_{\mathrm{1}}^{\mathrm{(i)}}=\kappa_{\mathrm{1}}$, $\kappa_{\mathrm{2}%
}^{\mathrm{(i)}}=\kappa_{\mathrm{2}}$. Neglecting all the fluctuations,
setting $da_{\text{\textrm{i}}}/dt=0$, and taking the expectation value with
respect to the steady state of Eq. (\ref{initial}), it is easy to find that%
\begin{equation}
\left(  i\Delta_{\mathrm{i}}-\frac{\kappa}{2}\right)  \left\langle
a_{\mathrm{i}}\right\rangle -\sqrt{\kappa_{\mathrm{1}}}\left\langle
a_{\mathrm{i}}^{\mathrm{in}}\right\rangle -\sqrt{\kappa_{\mathrm{2}}%
}\left\langle a_{\mathrm{i+1}}^{\mathrm{out}}\right\rangle =0, \label{second}%
\end{equation}
where $\kappa=\kappa_{\mathrm{0}}+\kappa_{\mathrm{1}}+\kappa_{\mathrm{2}}$
denotes the total dissipation of each cavity mode, and $\Delta_{\mathrm{i}}$
denotes the detuning $\omega-\omega_{\mathrm{i}}$. Also, using simple
recurrence relations, we obtain $\left\langle a_{\mathrm{i}}^{\mathrm{in}%
}\right\rangle =\left\langle a_{\mathrm{in}}\right\rangle +\sqrt
{\kappa_{\mathrm{1}}}\sum_{\mathrm{j=1}}^{\mathrm{i-1}}\left\langle
a_{\mathrm{j}}\right\rangle $ and $\left\langle a_{\mathrm{i}}^{\mathrm{out}%
}\right\rangle =\sqrt{\kappa_{\mathrm{2}}}\sum_{\mathrm{j=i}}^{\mathrm{N}%
}\left\langle a_{\mathrm{j}}\right\rangle $, where we have used $\left\langle
a_{\mathrm{1}}^{\mathrm{in}}\right\rangle =\left\langle a_{\mathrm{in}%
}\right\rangle $ and $\left\langle a_{\mathrm{N+1}}^{\mathrm{out}%
}\right\rangle =0$. Therefore, Eq. (\ref{second}) reduces to%
\begin{equation}
\left(  i\Delta_{\mathrm{i}}-\frac{\kappa}{2}\right)  \left\langle
a_{\mathrm{i}}\right\rangle -\sqrt{\kappa_{\mathrm{1}}}\left\langle
a_{\mathrm{in}}\right\rangle -\kappa_{\mathrm{1}}\sum_{\mathrm{j=1}%
}^{\mathrm{i-1}}\left\langle a_{\mathrm{j}}\right\rangle -\kappa_{\mathrm{2}%
}\sum_{\mathrm{j=i+1}}^{\mathrm{N}}\left\langle a_{\mathrm{j}}\right\rangle
=0. \label{third}%
\end{equation}
If $\kappa_{\mathrm{1}}=\kappa_{\mathrm{2}}$, Eq. (\ref{third}) can be further
simplified to%
\begin{equation}
\left\langle a_{\mathrm{i}}\right\rangle =\frac{\sqrt{\kappa_{\mathrm{1}}%
}\left\langle a_{\mathrm{in}}\right\rangle +\kappa_{\mathrm{1}}\sum
_{\mathrm{j=1}}^{\mathrm{N}}\left\langle a_{\mathrm{j}}\right\rangle }%
{i\Delta_{\mathrm{i}}-\kappa_{\mathrm{0}}/2}. \label{4}%
\end{equation}
Through simple algebra, we achieve%
\begin{equation}
\sum_{\mathrm{i=1}}^{\mathrm{N}}\left\langle a_{\mathrm{i}}\right\rangle
=\frac{\sqrt{\kappa_{\mathrm{1}}}c}{1-\kappa_{\mathrm{1}}c}\left\langle
a_{\mathrm{in}}\right\rangle , \label{5}%
\end{equation}
where the constant $c=\sum_{\mathrm{i=1}}^{\mathrm{N}}\left(  i\Delta
_{\mathrm{i}}-\kappa_{\mathrm{0}}/2\right)  ^{-1}$. According to the
output-input relations, the final output field which is our interest, can be
expressed as%
\begin{equation}
\left\langle a_{\mathrm{out}}\right\rangle =\left\langle a_{\mathrm{N+1}%
}^{\mathrm{in}}\right\rangle =\frac{1}{1-\kappa_{\mathrm{1}}c}\left\langle
a_{\mathrm{in}}\right\rangle \text{,} \label{6}%
\end{equation}
so the overall power transmission coefficient $\left\vert \mathcal{T}\left(
\omega\right)  \right\vert ^{2}=\left\vert \left\langle a_{\mathrm{out}%
}\right\rangle /\left\langle a_{\mathrm{in}}\right\rangle \right\vert ^{2}%
$\ for this system reads%
\begin{equation}
\left\vert \mathcal{T}\left(  \omega\right)  \right\vert ^{2}=\left\vert
1-\kappa_{\mathrm{1}}c\right\vert ^{-2}. \label{7}%
\end{equation}

Obviously, the transmission $\left\vert \mathcal{T}\left(  \omega\right)
\right\vert ^{2}$ has $N$ local minima $\left\vert \mathcal{T}\left(
\omega\right)  \right\vert _{\min}^{2}$ $\left(  \simeq0\right)  $ at
$\omega\simeq\omega_{\mathrm{i}}$ and $N-1$ local maxima $\left\vert
\mathcal{T}\left(  \omega\right)  \right\vert _{\max}^{2}$ at $\omega=\left(
\omega_{\mathrm{i}}+\omega_{\mathrm{i+1}}\right)  /2$. For the simplicity of
numerical simulation, we further assume that the resonant frequencies are
equally (or periodically) spaced between the adjacent cavities' modes, i.e.,
$\omega_{\mathrm{i+1}}-\omega_{\mathrm{i}}=\delta$.

The overall transmission is shown in the red solid lines of Fig. 2(a)-2(f),
which describe the cases of $N=1$, $2$, $3$, $4$, $5$, $6$, respectively. When
$N=1$, i.e., the system only includes one cavity, Fig. 2(a) actually depicts
the well-known transmission response curve of a single cavity mode
\cite{rokhsari}. When $N>1$, obviously, there exist some sharp peaks in the
middle of the adjacent cavities' modes, which is a direct analogy to the
phenomenon of electromagnetically induced transparency in atomic vapors
\cite{rmp}\ or semiconductors \cite{hailin}. These narrow peaks originate from
the interference effect of the cavities' delay. The output $a_{\mathrm{i+1}%
}^{\mathrm{out}}$ of the \textrm{i+1}-th cavity couples back into the
$\mathrm{i}$-th cavity as the second input port besides $a_{\mathrm{i}%
}^{\mathrm{in}}$.\textit{ }The transmission is nearly canceled when the light
is resonant with one of the cavities' modes due to the over-coupling regime
($\kappa_{\mathrm{1}},\kappa_{\mathrm{2}}\gg\kappa_{\mathrm{0}}$) between the
waveguides and resonators \cite{rokhsari}. However, in the middle the adjacent
modes, the destructive interference results in a very narrow transmission
resonance \cite{maleki}. We will discuss these in the following part.

For comparison, the blue dotted lines in Fig. 2(b)-2(f) describe the cases
that the output $a_{\mathrm{i+1}}^{\mathrm{out}}$ of the \textrm{i+1}-th
cavity directly decays into free space instead of coupling back into the
$\mathrm{i}$-th cavity. Therefore there is no interference effect between the
two input fields, and the whole output transmission describes the
collective\ response of all the cavities' modes.

\begin{figure}[ptb]
\centerline{\includegraphics[keepaspectratio=true,width=0.5\textwidth]{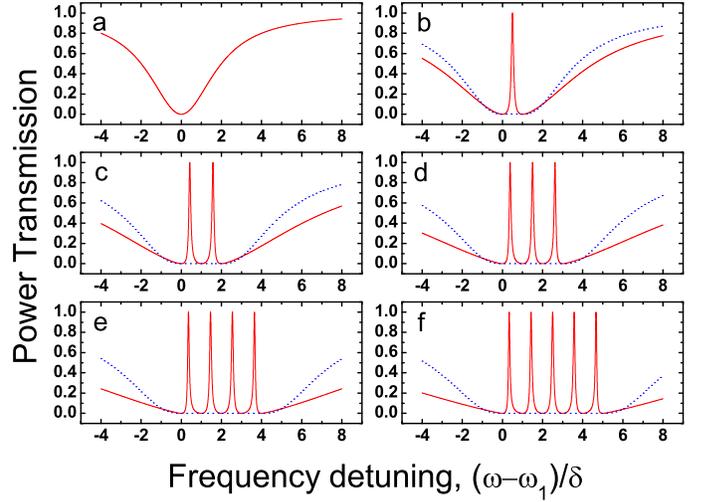}}\caption{(Color
online) Overall power transmission coefficient $\left\vert \mathcal{T}\left(
\omega\right)  \right\vert ^{2}$. Red solid (blue dotted) lines in Figure
2(a)-2(f) describe the system including one, two, three, four, five, six
cavities, respectively, and there are (no) side coupling among them. Other
parameters: $\kappa_{\mathrm{0}}/\kappa_{\mathrm{1}}=10^{-3}$, $\kappa
_{\mathrm{1}}/\delta=2$.}%
\end{figure}

To quantitatively characterize the above optical EIT, Fig. 3(a) and 3(b)
depict the full width at half maximum (FWHM) and maximal transmission rate of
the EIT windows respectively. As shown in Fig. 3(a), FWHM mostly depends on
the coupling strength $\kappa_{\mathrm{1}}$, that is, larger $\kappa
_{\mathrm{1}}$ leads to narrower peak for the given intrinsic cavity loss
$\kappa_{\mathrm{0}}$\ and the mode spacing $\delta$. In other words, once the
parameters $\kappa_{\mathrm{0}}$ and $\kappa_{\mathrm{1}}$ have been given,
the smaller $\delta$ causes sharper peak. It agrees with the experiental
prediction that there is no EIT phenomenon if $\kappa_{\mathrm{1}}\ll$
$\delta$, because at this time the spectrum differences between the adjacent
cavities' modes are so large that interference does not work significantly,
and the total transmission will represent the resonant absorptions of $N$
cavities' modes. Fig. 3(b) shows how the maximal transmission $\left\vert
\mathcal{T}\left(  \omega\right)  \right\vert _{\max}^{2}$ depends on the
intrinsic cavity loss $\kappa_{\mathrm{0}}$ for a given coupling strength
$\kappa_{\mathrm{1}}$. When the magnitude of $\kappa_{\mathrm{0}}$ is of the
order of$\ \delta$, $\left\vert \mathcal{T}\left(  \omega\right)  \right\vert
_{\max}^{2}$ will decrease rapidly to zero; when $\kappa_{\mathrm{0}}$ can be
arbitrarily small, $\left\vert \mathcal{T}\left(  \omega\right)  \right\vert
_{\max}^{2}$ almost maintains unitary since there is no external loss for the
optical resonator-waveguide system here.

\begin{figure}[ptb]
\centerline{\includegraphics[keepaspectratio=true,width=0.5\textwidth]{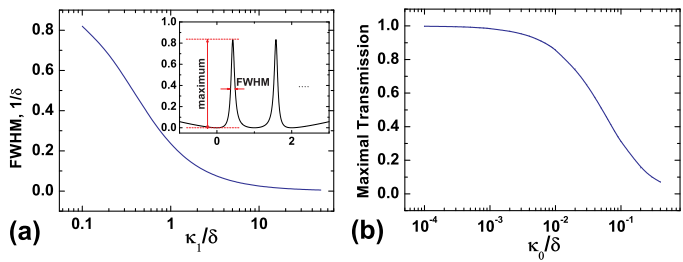}}\caption{(Color
online) (a) FWHM of the EIT windows vs. $\kappa_{\mathrm{1}}$. Other
parameter: $\kappa_{\mathrm{0}}/\delta=10^{-4}$. (b) Maximal power
transmission\ (peak value) $\left\vert \mathcal{T}\left(  \omega\right)
\right\vert _{\max}^{2}$\ vs. $\kappa_{\mathrm{0}}$. Other parameter
$\kappa_{\mathrm{1}}/\delta=2$.}%
\end{figure}

It is necessary to give some brief remarks on the analogy and difference
between the electromagnetically and coupled-resonator induced transparency. On
one hand, the coupled-resonator induced transparency can be described in the
language which is used in the thoroughly documented field of EIT, where the
role of the \textquotedblleft atom\textquotedblright\ is played by the cavity,
the \textquotedblleft atomic transition\textquotedblright\ is acted by the
cavity mode, and the "strong driving laser beam" is represented by the strong
coupling between the adjacent cavities. As shown in Fig. 4, we illustrate the
sketched level diagram of two coupled cavities, i.e., $N=2$. Due to the
over-coupling regime among the waveguides and cavities, a probe beam tuned on
near-resonance with the first cavity mode $a_{\mathrm{1}}$ will be strongly
coupled into the second waveguide. The application of the second cavity mode
$a_{\mathrm{2}}$\ and the strong coupling between $a_{\mathrm{1}}$ and
$a_{\mathrm{2}}$ will split the first cavity mode into two dressed modes
$a_{\mathrm{1}}^{\prime}$\ and $a_{\mathrm{1}}^{\prime\prime}$\ with different
energies. In this case, the input field with frequency $\omega_{\mathrm{1}}$
can enter into the second waveguide via two intermediate cavity modes,
$a_{\mathrm{1}}^{\prime}$\ and $a_{\mathrm{1}}^{\prime\prime}$, whose
detunings are of equal magnitudes and opposite signs. As a result, their
contributions to the second waveguide process cancel out in second-order
perturbation theory and the first cavity becomes transparent to the probe
beam. Remarkably, this is analogous to the conventional EIT in atomic medium.

On the other hand, unlike the conventional atomic EIT, the coupled-resonator
induced transparency needs the condition that the probe beam be tuned on
near-resonance but not exact-resonance with the cavity mode. This is because
the system works in the over-coupling regime which results in the light
transferring from waveguide 1 to 2 with near unit transfer efficiency. Thus
the power transmission from waveguide 1 to the second cavity is very small and
the returned power from the second to the first cavity is even smaller. As a
result, the transmission will be nearly canceled when the light is resonant
with the first cavity mode since the destructive interference does not play a
significant role in the light transmission. For more cavities, similar
treatments can be done, which lead to the above-mentioned multiple EIT-like
transmission spectrum.

\begin{figure}[ptb]
\centerline{\includegraphics[keepaspectratio=true,width=0.35\textwidth]{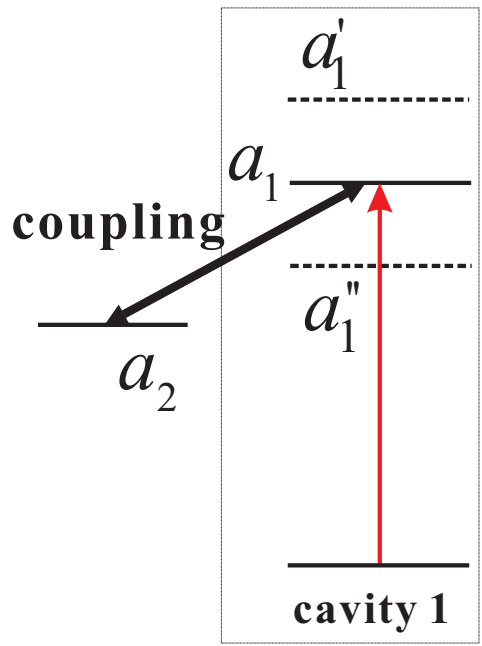}}\caption{(Color
online) Sketched level diagram of the first cavity coupled with the second.
Double-arrow bold line describes the strong coupling between the adjacent
cavities via two waveguides. $a_{\mathrm{1}}^{\prime}$\ and $a_{\mathrm{1}%
}^{\prime\prime}$\ are two dressed modes with different energies.}%
\end{figure}

We now turn to analyze the experimental feasibility of the present multiple
optical EIT. For this we briefly discuss two systems, microrings
\cite{lipson,xu} and microdisks \cite{kipp}. Recently, such cavities have been
considered as candidates for quantum information processing
\cite{buck,brun,xiao,shen,waks}. For microring resonators, the adjacent rings
can be linked by two etched waveguides, so the resonators (microrings) and
data channels (waveguides) can be integrated on a chip. Modes in microrings
posses high quality factors (up to $1.4\times10^{5}$ \cite{niehusmann}) and
ultra-small mode volumes. The current silicon-on-insulator (SOI) technology
allows for a high etching precision\ (better than $0.02$ $%
%TCIMACRO{\unit{\U{3bc}m}}%
%BeginExpansion
\operatorname{\mu m}%
%EndExpansion
$) by designing the mask and controlling inductively-coupled-plasma reactive
ion etching, so that the resonant detuning $\delta$ of the microrings and the
condition $\exp\left(  i\phi\right)  =1$ are easily satisfied. For microdisk
resonators, quality factors in excess of 1 million have been demonstrated in
micron scale silicon nitride (SiN$_{x}$) \cite{painter} and silica \cite{kipp}
at near visible wavelengths and in the $1550$-$%
%TCIMACRO{\unit{nm}}%
%BeginExpansion
\operatorname{nm}%
%EndExpansion
$ band, respectively. In-and-out coupling can be achieved by use of two fiber
tapers \cite{rokhsari}. To achieve\ and modulate the resonant detuning
$\delta$, the microdisk modes can be first positioned with an accuracy of
$0.5$ $%
%TCIMACRO{\unit{nm}}%
%BeginExpansion
\operatorname{nm}%
%EndExpansion
$ using standard lithographic techniques. For a more accurate modulation, wet
chemical etching introduced in References \cite{xudong,painter} (better than
$0.2$ $%
%TCIMACRO{\unit{pm}}%
%BeginExpansion
\operatorname{pm}%
%EndExpansion
$) and temperature tuning discussed in Reference \cite{temperature} can be applied.

In conclusion, we theoretically present an all-optical scheme to directly
simulate multiple EIT using periodically side-coupled $N$ resonators. The
highly sharp EIT-like transparency windows only rely on the frequencies of the
cavities' modes, so that it is selective through the design of resonator
array. This is of importance for applications in optical communications (e.g.,
channel-selective bandpass filters) and quantum information processing (e.g.,
slow light systems).

\begin{acknowledgments}
This work was supported by National Fundamental Research Program, also by
National Natural Science Foundation of China (Grant No. 10674128 and 60121503)
and the Innovation Funds and \textquotedblleft Hundreds of
Talents\textquotedblright\ program of Chinese Academy of Sciences and Doctor
Foundation of Education Ministry of China (Grant No. 20060358043).
\end{acknowledgments}

\end{document}